\documentclass[journal]{IEEEtran}
\normalsize
\usepackage{amssymb,amsthm}
\usepackage{psfrag}

\usepackage{cite}

\ifCLASSINFOpdf
   \usepackage[pdftex]{graphicx}
\else
   \usepackage[dvips]{graphicx}
\fi
\usepackage[cmex10]{amsmath}
\usepackage{algorithm}
\usepackage{algpseudocode}
\usepackage{fixltx2e}

\usepackage{stfloats}
\begin{document}
%
\title{Access Point Density and Bandwidth Partitioning in Ultra Dense Wireless Networks}
%
%
%

\author{Stelios~Stefanatos~and~Angeliki~Alexiou,~\IEEEmembership{Member,~IEEE}
\thanks{The authors are with the Department
of Digital Systems, University of Piraeus, Greece.

This work has been performed in the context of THALES-INTENTION (MIS: 379489) research project, within the framework of Operational Program ``Education and Lifelong earning'', co-financed by the European Social Fund (ESF) and the Greek State.}
}

%
%

\markboth{}%
{}
%



\maketitle

\begin{abstract}
This paper examines the impact of system parameters such as access point density and bandwidth partitioning on the performance of randomly deployed, interference-limited, dense wireless networks. While much progress has been achieved in analyzing randomly deployed networks via tools from stochastic geometry, most existing works either assume a very large user density compared to that of access points, which does not hold in a dense network, and/or consider only the user signal-to-interference-ratio as the system figure of merit, which provides only partial insight on user rate as the effect of multiple access is ignored. In this paper, the user rate distribution is obtained analytically, taking into account the effects of multiple access as well as the SIR outage. It is shown that user rate outage probability is dependent on the number of bandwidth partitions (subchannels) and the way they are utilized by the multiple access scheme. The optimal number of partitions is lower bounded for the case of large access point density. In addition, an upper bound of the minimum access point density required to provide an asymptotically small rate outage probability is provided in closed form.

\end{abstract}


%
\IEEEpeerreviewmaketitle

\section{Introduction}
%
%
%
%
\IEEEPARstart{S}{mall} cell networks have attracted a lot of attention recently as they are considered a promising method to satisfy the ever increasing rate demands of wireless users. Some studies have suggested that, by employing low cost access points (APs), the density $\lambda_a$ of APs will potentially reach, or even exceed, the density $\lambda_u$ of user equipments (UEs), therefore introducing the notion of ultra dense wireless networks \cite{Qualcomm}. With a large number of APs available, a random UE will most probably connect to a strong signal AP, having to share the AP resources with a limited number of co-served UEs, and, ultimately, achieve high rates. However, in order to exploit the full system resources, a universal frequency reuse scheme is employed which inevitably results in significant interference that has to be taken carefully into account in system design and performance analysis.

\subsection{Related Works and Motivation}

With an increasing network density, the task of optimally placing the APs in the Euclidean plane becomes difficult, if not impossible. Therefore, the APs will typically have an irregular, random deployment, which is expected to affect system performance. Recent research has showed that such randomly deployed cellular systems can be successfully analyzed by employing tools from stochastic geometry \cite{Baccelli, ElSawy}. While significant results have been achieved, most of these works assume $\lambda_u \gg \lambda_a$, effectively ignoring UE distribution, and/or  consider only the user signal-to-interference ratio (SIR). Assumption $\lambda_u \gg \lambda_a$ does not hold in the case of dense networks, whereas SIR provides only partial insight on the achieved user rate as the effect of multiple access is neglected \cite{AndrewsMag}.

A few recent works have attempted to address these issues. Specifically, the UE distribution is taken into account in \cite{Lee,Li} by incorporating in the analysis the probability of an AP being inactive (no UE present within its cell). However, analysis considers only the SIR. In \cite{Cao,Singh} the UE distribution is employed for computation of user rates under time-division-multiple-access (TDMA) without considering the effect of SIR outage. In addition, TDMA may not be the best multiple access scheme under certain scenarios.

Partitioning the available bandwidth and transmitting on one of the resulting subchannels (SCs), i.e., frequency-division-multiple-access (FDMA), has been shown in \cite{Andrews_bandwidth} to be beneficial for the case of ad-hoc networks assuming a channel access scheme where each node transmits independently on a randomly selected SC. This decentralized scheme was employed in \cite{Huang} for modeling the uplink of a cellular network with  frequency hopping channel access. However, this approach is  inappropriate for a practical cellular network where scheduling decisions are made by the AP and transmissions are orthogonalized to eliminate intra-cell interference (no sophisticated processing at receivers is assumed that would allow for non-orthogonal transmissions). A straightforward modification of the bandwidth partitioning concept for the downlink cellular network was considered in \cite{Andrews} where UEs are multiplexed via TDMA and transmission is performed on one, randomly selected SC. This simple scheme was shown to provide improved SIR performance, however, with no explicit indication of how many partitions should be employed or how performance would change by allowing more that one UEs transmitting at the same time slot on different SCs.

\subsection{Contributions and Paper Organization}

In this paper, the stochastic geometry framework is employed for analyzing the downlink user rate of a dense wireless network under a multiple access scheme that exploits bandwidth partitioning for \emph{both} interference reduction and efficient resource sharing among UEs. The previously mentioned issues are explicitly addressed by considering in the analysis
\begin{itemize}
\item{
    the UE distribution,
    }
\item{
    a multiple access scheme that allows for parallel orthogonal transmissions in frequency,
    }
\item{
    the effect of SIR outage.
    }
\end{itemize}
Under this framework, the user rate distribution is analytically derived for two instances of multiple access schemes that reveals dependence of performance on the number of bandwidth partitions as well as the way the are utilized. The analytical rate distribution expression, apart from allowing for efficient numerical optimization of system parameters, is employed to derive a  closed-form lower bound of the optimal number of partitions for the case of large AP density, as well as a closed-form upper bound of the minimum AP density required to provide a given, asymptotically small, rate outage probability. The latter is of critical importance given the trend of AP densification in future wireless networks. Numerical results demonstrate the merits of increased AP density, as well as efficient use of bandwidth partitions, in enhancing network performance in terms of achieved user rate.

The paper is organized as follows. Section II describes the system model, along with a discussion on the suitability of various metrics with respect to (w.r.t.) UE performance. In Section III, the user rate distribution is analytically obtained for two instances of multiple access schemes. The number of SCs that minimizes rate outage probability is investigated in Section IV, and Section V  provides a closed form upper bound of the minimum required AP density that can support a given, asymptotically small, user rate outage probability. Section VI presents numerical examples that provide insights on various system design aspects, and Section VII concludes the paper.

\section{System Model and Performance Metrics}
The downlink of an interference-limited, dense wireless network is considered. Randomly deployed over $\mathbb{R}^2$ APs and UEs are modeled as independent homogeneous Poisson point processes (PPP) $\Phi_a$, $\Phi_u$, with densities $\lambda_a$, $\lambda_u$, respectively. Full buffer transmissions and Gaussian signaling are assumed, with interference treated as noise at the receivers. Each UE is served by its closest AP resulting in irregular, disjoint cell shapes forming a Voronoi tessellation of the plane \cite{Baccelli}. Elimination of intra-cell interference is achieved by an orthogonal FDMA/TDMA scheme with the total system bandwidth partitioned offline to $N$ equal size SCs. All active APs in the system transmit at the same power over all (active) SCs, with the power selected appropriately large so that the system operates in the interference limited region in order to maximize spectral efficiency \cite{Lozano}. No coordination among APs is assumed, i.e., each AP makes independent scheduling decisions.

Considering a typical UE located at the origin and served by its closest AP of index, say, $0$, the SIR achieved at SC $n \in \{1, 2, \ldots, N\}$ is given by
\begin{equation} \label{eq:1}
\textrm{SIR}_{n} = \frac{g_{0,n} r_0^{-\alpha}}{\sum_{i\in \Phi_{a}\setminus \{0\}}\delta_{i,n} g_{i,n}r_i^{-\alpha}},
\end{equation}
where $r_0\geq0$ is the distance from the serving AP, $\alpha > 2$ the path loss exponent, and $g_{0,n}\geq0$ an exponentially distributed random variable with unit mean, modeling small scale (Rayleigh) fading. The denominator in (\ref{eq:1}) represents the interference power at the considered SC, where $g_{i,n}$, $r_i$ are the channel fading and distance of  AP $i$ w.r.t. the typical UE, respectively, and $\delta_{i,n} \in \{0,1\}$ is an indicator variable representing whether AP $i$ transmits on SC $n$. Note that $\delta_{i,n}$ depends on the total number of UEs associated with AP $i$ as well as the multiple access scheme and its presence in (\ref{eq:1}) is to account for APs that do not interfere due to lack of associated UEs and/or scheduling decisions. Channel fadings $\{g_{i,n}\}$ are assumed independent, identically distributed (i.i.d.) w.r.t. AP index $i$.

$\textrm{SIR}_{n}$ is a random variable due to the randomness of fading, AP and UE locations, as well as the multiple access scheme, and its statistical characterization is of interest. To this end, the interference term of (\ref{eq:1}) can be viewed as shot-noise generated by a marked PPP \cite{Baccelli} of density $\lambda_a$ outside a ball of radius $r_0$ centered at the origin, and marks $\{g_{i,n},\delta_{i,n}\}$. Statistical characterization of a marked PPP can be obtained by standard methods when the following conditions hold \cite{Baccelli}:
\begin{enumerate}
\item{marks are mutually independent given the location of points, and,}
\item{each mark depends only on the location of its corresponding point.}
\end{enumerate}
Channel fadings $\{g_{i,n}\}$ satisfy both conditions by assumption, whereas variables $\{\delta_{i,n}\}$ satisfy only the first due to their dependence on the total number of UEs associated with each AP. By fundamental properties of the PPP, the numbers of UEs associated with different APs are mutually independent since AP cells are disjoint. For each AP, the number of associated UEs is determined by $\lambda_u$ and its cell area, with the latter depending not only on its own position but also on the position of its neighbours APs as well, rendering condition (2) invalid for $\{\delta_{i,n}\}$. In order to obtain tractable expressions for the SIR distribution the following assumption is adopted:
\newtheorem{assumption}{Assumption}
\begin{assumption}
The number $K$ of UEs associated with a random AP is independent of $\Phi_a$.
\end{assumption}
Note that this assumption is actually stronger than  the second condition but is convenient as it allows for incorporating the averaged-over-$\Phi_a$ probability mass function (PMF) of $K$ that will be used later in the analysis, given by the following lemma:
\newtheorem{lemma}{Lemma}
\begin{lemma}
The PMF of the number $K$ of UEs associated with a randomly chosen AP, averaged over the statistics of $\Phi_a$, is \cite{Yu}
\begin{equation} \label{eq:Kpmf}
\Pr\{K\}=\frac{3.5^{3.5}\Gamma(K+3.5)\tau^{3.5}}{\Gamma(3.5)K!(1+3.5\tau)^{K+3.5}},  K \geq0,
\end{equation}
where $\Gamma(\cdot)$ is the Gamma function and $\tau \triangleq \lambda_a/\lambda_u$.
\end{lemma}
Note that $\Pr\{K\}$ is a decreasing function of $\tau$ and the mean of $K$ equals $1/\tau$. The above approach was shown in \cite{Lee,Li} to provide accurate results and will be validated by simulations in Sect. III.

Under Assumption 1, $\{\delta_{i,n}\}$ are i.i.d. over $i$ and characterized by the \emph{activity probability} $p_n \triangleq \Pr\{\delta_{i,n}=1\} \in (0,1], \forall i$, whose actual value will be investigated in Sect. III for specific multiple access schemes. The cumulative distribution function (CDF) of $\textrm{SIR}_{n}$ can now be obtained in a simple expression as given by the following lemma:

\newtheorem{lemma2}{Lemma}
\begin{lemma}
The CDF of $\textrm{SIR}_{n}$ under an activity probability $p_{n}$ is given by \cite{Andrews}
\begin{equation} \label{eq:FSIR}
F_{\textrm{SIR}_{n}}(\theta) \triangleq \Pr\{\textrm{SIR}_{n} \leq \theta\} = 1-\frac{1}{1+p_{n}\rho(\theta)},
\end{equation}
for $\theta \geq 0$, where $\rho(\theta) \triangleq \theta^{2/\alpha}\int_{\theta^{-2/\alpha}}^{\infty}1/\left(1+u^{\alpha/2}\right)du$.
\end{lemma}
Note that setting $p_{n}$ \emph{a-priori} equal to 1, as in, e.g., \cite{Cao,Singh}, implies that there is always a UE available to be allocated in every AP of the system, i.e., $\lambda_u  \gg \lambda_a$, which is not the case in dense networks. For example, for the case $\lambda_a=\lambda_u$ and noting that $p_n \leq \Pr\{K>0\}$ for any multiple access scheme, it follows from (\ref{eq:Kpmf}) that $p_n \leq 0.58$.

Knowledge of (\ref{eq:FSIR}) is of importance as it provides the probability $F_{\textrm{SIR}_{n}}(\theta_0)$ of service outage due to inability of UE  operation below SIR threshold $\theta_0$ whose value may be dictated by operational requirements, e.g., synchronization, and/or application (QoS) requirements. In addition, $F_{\textrm{SIR}_{n}}(\theta)$ can be used to obtain CDFs of other directly related quantities of interest by transformation of variables. One such quantity employed extensively in the related literature, e.g., \cite{Andrews,Li}, is the rate $\overline{R}_{n}$ achieved \emph{per channel use}, i.e, on a single time slot, given by
\begin{equation} \label{eq:R_bound}
\overline{R}_{n} \triangleq \frac{1}{N} \log_2(1+\textrm{SIR}_{n}) \textrm{ (b/s/Hz).}
\end{equation}
Examining $\overline{R}_{n}$ is important from the viewpoint of system throughput \cite{Li} but provides little insight on the achieved user rate. Note that $\overline{R}_{n}$ is an upper bound on the actual user rate. In case when the considered SC has to be time-shared among UEs, rate will only be a fraction of $\overline{R}_n$. In an attempt to remedy this issue, $\overline{R}_n$ was divided by the (random) number of UEs sharing the SC in \cite{Cao,Singh} (case of $N=1$ was only considered). However, this is still a misleading measure of performance as it does not take into account the probability of an SIR outage and, therefore, provides overconfident results.

In order to avoid these issues, the achieved rate of a typical UE is defined in this paper as
\begin{equation} \label{eq:4}
R_{n}\triangleq \mathbf{1}\{\textrm{SIR}_{n} \geq \theta_0\} \frac{\overline{R}_{n}}{(L_{n}+1)}  \textrm{ (b/s/Hz)},
\end{equation}
where $\mathbf{1}\{\cdot\}$ is the indicator function and integer $L_{n} \geq 0$ is the number of time slots between two successive transmissions to the typical UE, referred to as \emph{delay} in the following. Clearly, (\ref{eq:4}) takes into account both the effects of multiple access and SIR outage via $L_{n}$ and the indicator function, respectively. In order to obtain the rate outage probability, i.e., the CDF of $R_n$, (statistical) evaluation of $p_n$ and $L_n$ is required, both depending, in addition to the UE distribution, on the multiple access scheme that is investigated in the following section.

\section{Effect of Multiple Access on Achievable User Rates}

In this section the effect of multiple access on the achievable user rate is investigated. The multiple access scheme must strive to maximize UE resource utilization, while at the same time minimize inter-cell interference. These are conflicting requirements which, as it will be shown, can be (optimally) balanced by the choice of $N$. For analytical purposes the two schemes considered below are non-channel aware, with the corresponding performance serving as a lower bound under a channel aware resource assignment scheme, and fair, i.e., there are no priorities among UEs.

\subsection{TDMA}
The following scheme, referred to in the following as TDMA, will serve as a baseline.

\begin{algorithm}
\caption{TDMA}\label{MAC}
\begin{algorithmic}[1]
\Statex - UEs are multiplexed via TDMA.
\Statex - Transmission to any UE is performed on one, randomly selected SC out of total $N$.
\end{algorithmic}
\end{algorithm}

This scheme is a straightforward application of the random SC selection scheme employed in adhoc studies \cite{Andrews_bandwidth} to the downlink cellular setting, and can be also  viewed as a generalization of conventional TDMA ($N=1$) that is usually assumed in works on cellular networks. It was first examined in \cite{Andrews}, where it was shown that it provides improved SIR coverage by using essentially the same principle as in a frequency hopping scheme.

\subsection{FDMA/TDMA}

The major argument against TDMA is the inability of parallel transmissions in frequency by multiple UEs when $N>1$, which is expected to be beneficial under certain operational scenarios. In this paper, the following simple modification is employed, referred to as FDMA/TDMA in the following, that allows for multiple UEs served at a single time slot ($\lfloor \cdot \rfloor$ denotes the largest smallest integer operator).

\begin{algorithm}
\caption{FDMA/TDMA}
\begin{algorithmic}[1]
\Statex - Define $\mathcal{N} \subseteq \{1,2,\ldots,N\}$ the set of available SCs for allocation at any given instant.
\Statex - Randomly order the $K$ cell users via an index $k \in \{1,2,\ldots,K\}$.
\For{$L=0$ to $\lfloor K/N \rfloor$} 
\State $\mathcal{N} \gets \{1,2,\ldots,N\}$;
\For{$k=LN+1$ to $\min\{LN+N,K\}$} 
\State Assign UE $k$ a random SC $n_k \in \mathcal{N}$; 
\State $\mathcal{N} \gets \mathcal{N}\backslash\{n_k\}$;
\EndFor
\EndFor
\Statex - UEs sharing the same SC are multiplexed via TDMA.
\end{algorithmic}
\end{algorithm}

Note that the above scheme also subsumes conventional TDMA as a special case. Two typical realizations of the scheme for $N=3$ are shown in Fig. 1. As can be seen, there will be cases with unused SCs ($K<N$) or SCs that support one additional UE compared to others ($K>N$) with no action taken to compensate for these effects. The inefficient bandwidth utilization is not a real issue since presence of unused SCs is beneficial in terms of reduced interference and also $N$ is variable that can be set to a small enough value so that this event is avoided, if desired. The load imbalance among SCs is irrelevant for rate computations due to averaging.

Having specified the multiple access schemes, the corresponding quantities $p_{n}$ and $L_{n}$ will be evaluated in the following subsections. By the symmetry of the system model and the schemes considered, all the performance metrics presented in Sect. II do not depend on $n$. Therefore, the typical UE will be considered assigned to SC 1 and index $n$ will be dropped from notation in the following.

\begin{figure}
\centering
\resizebox{\columnwidth}{!}{\includegraphics{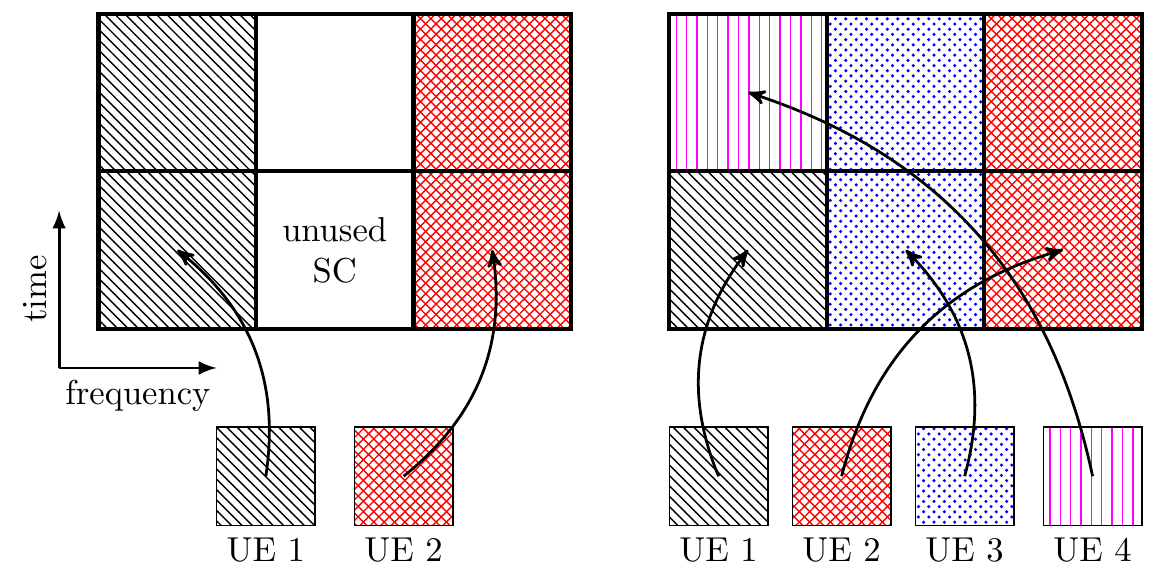}}
\caption{Typical realizations of the FDMA/TDMA scheme, $K<N$ (left) and $K>N$ (right).}
\end{figure}

\subsection{Computation of Activity Probability}
The following lemma holds for the activity probability $p$ of TDMA and FDMA/TDMA.
\newtheorem{lemma4}{Lemma}
\begin{lemma}
Under assumption 1, the activity probability of any AP in the system, other than $0$, is
\begin{equation} \label{eq:activ_prob}
p =  \left\{\begin{IEEEeqnarraybox}[\relax][c]{l's}
\frac{1}{N}\Pr\{K>0\},& for TDMA,\\
\frac{1}{N}\sum_{K>0}\Pr\{K\}\min\{K,N\},& for FDMA/TDMA,
\end{IEEEeqnarraybox}\right.
\end{equation}
with $\Pr\{K\}$ as given in Lemma 1.
\end{lemma}
\begin{IEEEproof}
See Appendix A.
\end{IEEEproof}

As expected, $p$ is a decreasing function of $N$ in both cases, and it can be easily shown that, for the same $N>1$, $p$ of FDMA/TDMA is lower bounded by the corresponding $p$ of TDMA with equality when $\Pr\{K \leq 1\} = 1$, i.e., with a dense AP deployment. Note that in \cite{Andrews}, $p$ for TDMA was set equal to $1/N$, implying that $\Pr\{K>0\}=1$, which is (approximately) valid only for $\tau \rightarrow 0$. Substituting (\ref{eq:activ_prob}) in (\ref{eq:FSIR}) shows that $F_{\textrm{SIR}}(\theta)$ is decreasing in $N$, i.e., bandwidth partitioning improves performance in terms of SIR. 

Figure 2 shows the behavior of $p$ as a function of $N$ for FDMA/TDMA and TDMA and various values of $\tau$. Conventional TDMA performance corresponds to $N=1$. It can be seen that both schemes outperform conventional TDMA, with larger values of $N$ required to obtain the same $p$ under heavier system load. TDMA is always better than FDMA/TDMA, especially under heavy system load. For $\tau=10$, both schemes essentially operate exactly the same and this is reflected on the values of $p$. For $\tau=1$, FDMA/TDMA is worse than TDMA but relatively close, with similar dependence on $N$ (inversely proportional).

\emph{Remark}: According to the previous discussion, the SIR grows unbounded with increasing $\tau$ and/or $N$, which is unrealistic. However, arbitrarily large values of $\tau$ are not of interest due to practical considerations, whereas arbitrarily large values of $N$ are not acceptable from a user rate perspective as will be shown in Sect. IV (Lemma 6).

\begin{figure}
\centering
\resizebox{\columnwidth}{!}{\includegraphics{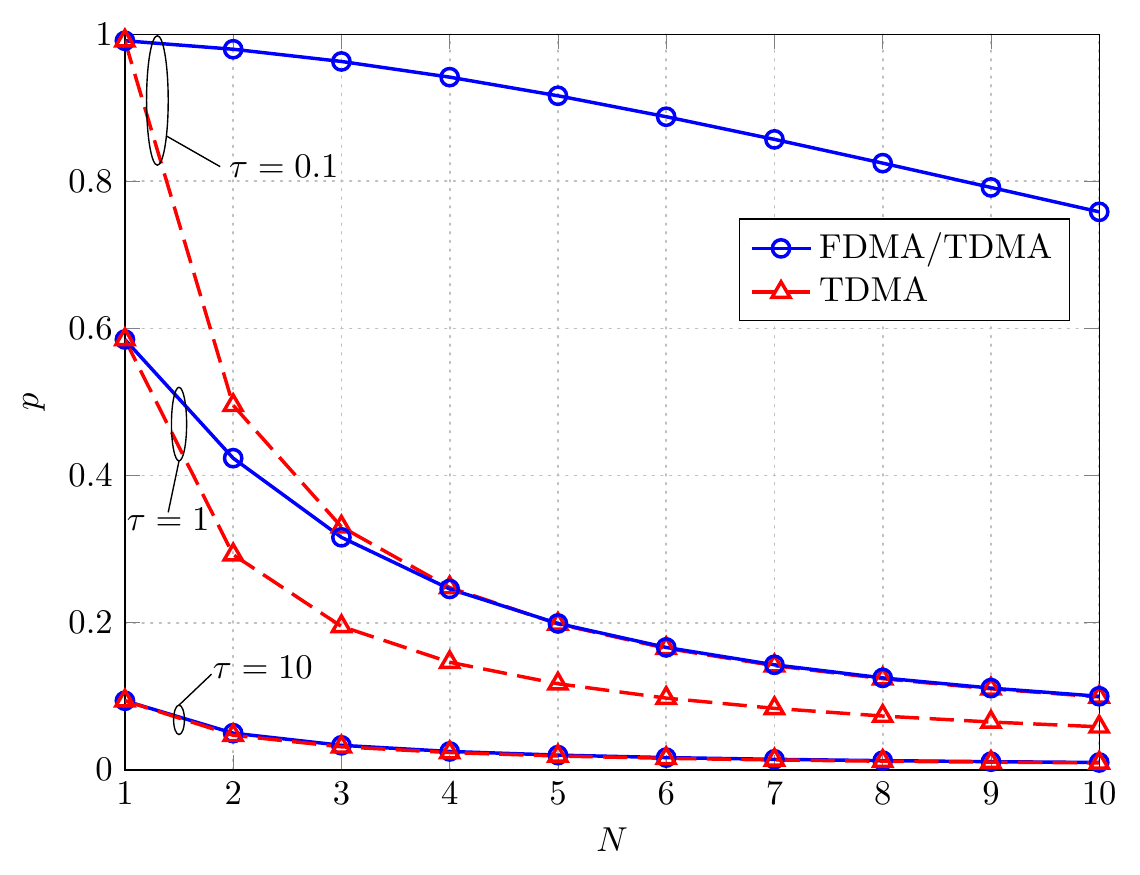}}
\caption{Activity probability $p$ for FDMA/TDMA and TDMA.}
\end{figure}

\subsection{Computation of Delay}
Computation of delay requires knowledge of the distribution of the total number $K_0$ of UEs associated with AP 0, \emph{in addition to} the typical UE. The PMF of (\ref{eq:Kpmf}) does not hold for AP 0 as conditioning on its area covering the position of the typical UE makes it larger than the cell area of a random AP \cite{Baccelli}. Taking this fact into account, the PMF of $K_0$ can be shown to be given as in the following lemma.
\newtheorem{lemma3}{Lemma}
\begin{lemma}
The PMF of the number $K_0$ of UEs associated with AP 0, in addition to the typical UE, is \cite{Yu}
\begin{equation} \label{eq:K0pmf}
\Pr\{K_0\}=\frac{3.5^{4.5}\Gamma(K_0+4.5)\tau^{4.5}}{\Gamma(4.5)K!(1+3.5\tau)^{K_0+4.5}},  K_0 \geq0.
\end{equation}
\end{lemma}
For the case of TDMA, it is clear that $L=K_0$, whereas $L$ for FDMA/TDMA is given in the following lemma.
\newtheorem{lemma5}{Lemma}
\begin{lemma}
Define the event $\mathcal{A}_l \triangleq$ \{$l$ UEs assigned on SC 1 in addition to the typical UE\}, $l\geq0$. The PMF of $L$ for FDMA/TDMA equals
\begin{equation} \label{eq:p_L}
\Pr\{L\}=\sum_{K_0\geq0}\Pr\{K_0\}\Pr\{\mathcal{A}_{L} | K_0\},
\end{equation}
with $\Pr\{K_0\}$ as given in Lemma 4,
\begin{equation} \label{eq:p_L0_K}
\setlength{\nulldelimiterspace}{0pt}
\Pr\{\mathcal{A}_0 | K_0\}=\left\{\begin{IEEEeqnarraybox}[\relax][c]{l's}
1,& $0 \leq K_0 \leq N-1$, \\
\frac{2N-K_0-1}{N}, &$N \leq K_0 \leq 2N-2 $,\\
0,&$K_0 \geq 2N-1$,%
\end{IEEEeqnarraybox}\right.
\end{equation}
and
\begin{equation}
\Pr\{\mathcal{A}_l | K_0\}=\begin{cases}
0,& \textrm{$0 \leq K_0 \leq l N-1$}, \\
\frac{K_0-lN+1}{N},& \textrm{$lN \leq K_0 \leq (l+1)N-1$,}\\
\frac{(l+2)N-K_0-1}{N},&\parbox[t]{1\textwidth}{$(l+1)N \leq K_0$\\ \textrm{\phantom{xxxxx} $\leq (l+2)N-2$,}}\\
0,& \textrm{$K_0 \geq (l+2)N-1$},
\end{cases}
\end{equation}


for $l \geq 1$. 

\end{lemma}
\begin{IEEEproof}
Follows by the same arguments as in the proof of Lemma 3.
\end{IEEEproof}

Figure 3 shows the mean value of $L$ as a function of $N$ for FDMA/TDMA and TDMA and various values of $\tau$. Both schemes provide reduced delay by increasing $\tau$, since the number of UEs associated with the AP is reduced. For the case of FDMA/TDMA, average $L$ decreases also with $N$ as there are more SCs available to UEs and the probability of time sharing one of them by many UEs is reduced. On the other hand, for TDMA, $L$ is independent of $N$ since the availability of SCs is not exploited for parallel transmissions.

\begin{figure}
\centering
\resizebox{\columnwidth}{!}{\includegraphics{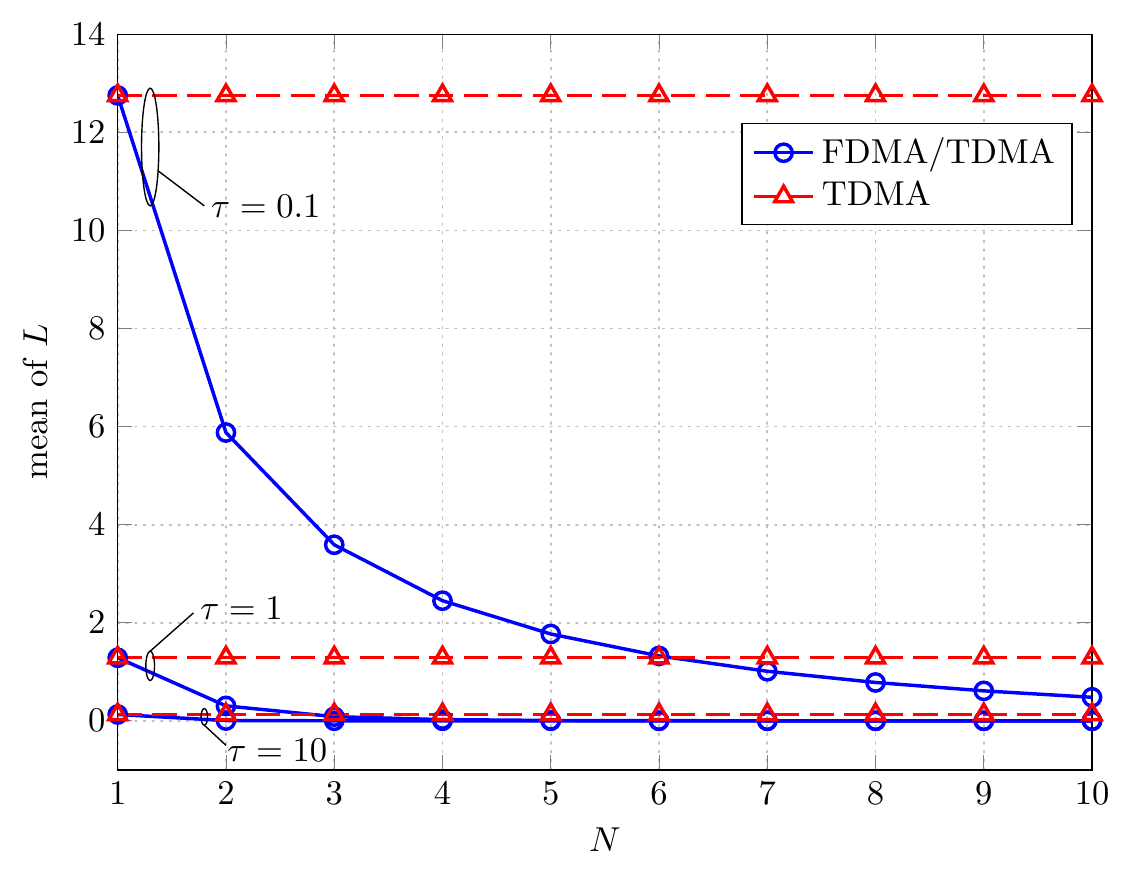}}
\caption{Mean value of delay $L$ for FDMA/TDMA and TDMA.}
\end{figure}

\subsection{Computation of Rate}
Obviously, FDMA/TDMA is advantageous when delay is considered, whereas, TDMA is more robust to interference. However, it is the achieved UE rate that is of more interest and, at this point, there is no clear indication which of the two schemes is preferable under this performance metric, i.e., what is of more importance, robustness to interference or efficient multiple access. Having specified the statistics of $p$ and $L$, the CDF of $R$ can now be obtained for both multiple access schemes as follows.

\newtheorem{proposition}{Proposition}
\begin{proposition}
The CDF of $R$ equals
\begin{equation} \label{eq:F_R}
F_{R}(r) = \sum_{L\geq0} \Pr\{L\} F_{R}(r|L),
\end{equation}
with $\Pr\{L\}$ as given in Sect. III. D and $F_{R}(r|L)$ the CDF of $R$ conditioned on the value of $L$, given by
\begin{equation} \label{eq:F_R_L}
\setlength{\nulldelimiterspace}{0pt}
F_{R}(r|L)=\left\{\begin{IEEEeqnarraybox}[\relax][c]{l's}
 F_{\textrm{SIR}}(\theta_0),& $r \leq \frac{\overline{R}_{\theta_0}}{N(L+1)}$, \\
F_{\textrm{SIR}}\left(2^{rN(L+1)}-1\right),& $r \geq \frac{\overline{R}_{\theta_0}}{N(L+1)}$,
\end{IEEEeqnarraybox}\right.
\end{equation}
where ${\overline{R}_{\theta_0}} \triangleq \log_2(1+\theta_0)$ is the minimum achievable rate for $N=1$ and $K_0=0$ (no contending UEs), conditioned on UE operation above SIR threshold $\theta_0$. 
\end{proposition}
\begin{IEEEproof}
See Appendix B.
\end{IEEEproof}

Note that the upper term of (\ref{eq:F_R_L}) indicates that for small values of $r$, rate outage probability coincides with the SIR outage probability, irrespective of the actual value of $r$, as for this rate region it is the SIR outage event      (strong interference) that prevents UEs from achieving these rates. For larger rates, $L$ appears in the lower term of (\ref{eq:F_R_L}), i.e., multiple access also affects performance in addition to interference.


\begin{figure}
\centering
\resizebox{\columnwidth}{!}{\includegraphics{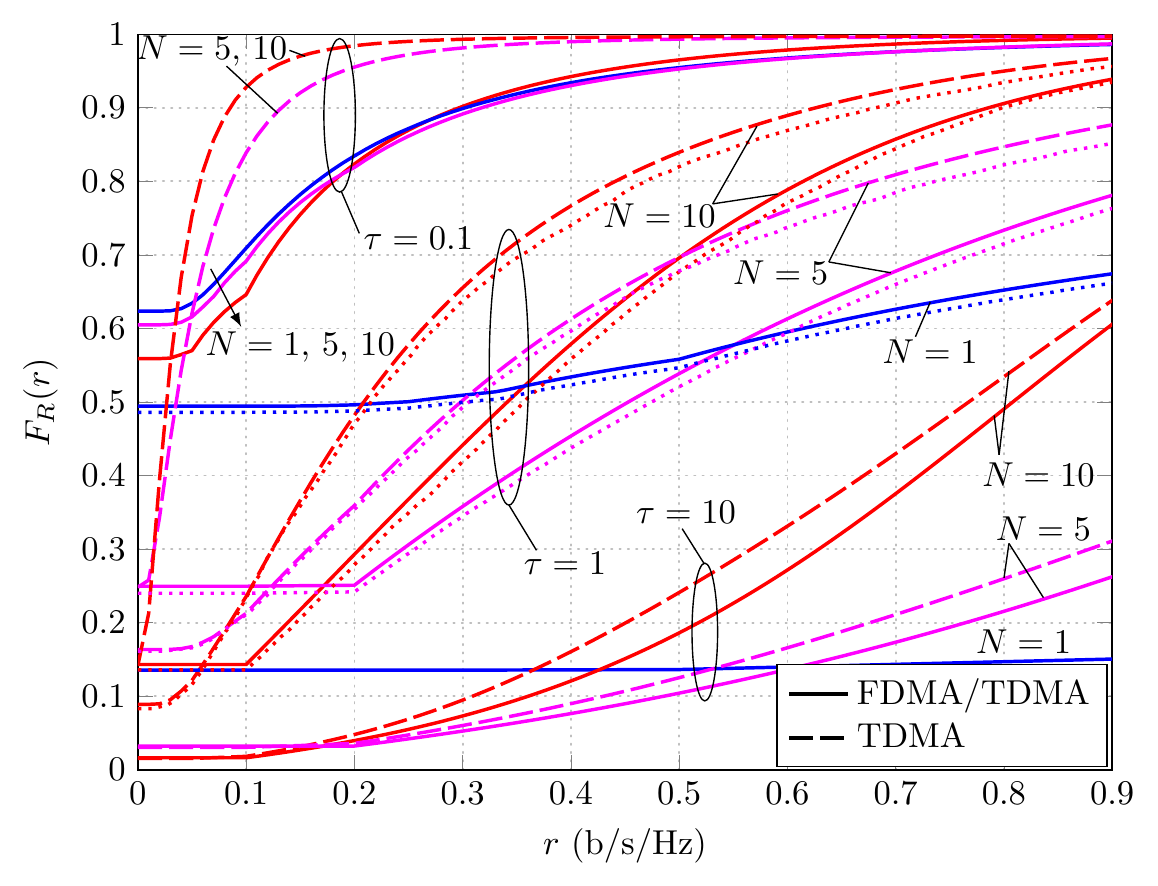}}
\caption{$F_{R}(r)$ of FDMA/TDMA and TDMA for various values of $N$ and $\tau$ ($\theta_0 =0 \textrm{ dB}$, $\alpha = 3$). Dotted lines depict simulation results.}
\end{figure}

Figure 4 shows $F_{R}(r)$ for $\theta_0 = 0 \textrm{ dB}$, $\alpha = 3$, $N = 1, 5, 10$, and various values of $\tau$, for FDMA/TDMA and TDMA. For the case of $\tau=1$ each analytical CDF is accompanied by the corresponding empirical CDF (dotted lines) obtained by simulations (simulation results for other $\tau$ values are omitted for clarity). The  good match between analysis and simulation validates the use of the derived formulas for system analysis and design.

As it can be seen, increasing AP density, i.e., increasing $\tau$, results in improved performance as the distance between UE and serving AP, as well as the number of UEs sharing the resources of a single AP, are reduced, which overbalance the effect of reduced distance from interfering APs. Concerning the dependence of rate on $N$, it can be seen that setting $N=1$ (conventional TDMA) is optimal when large data rates are considered, irrespective of $\tau$. However, the shape of the CDF for $N=1$ indicates a highly unfair system. Increasing $N$ results in a progressively more fair system, favoring the small-rate operational region. In particular, for $\tau=1$, and assuming a rate outage when the typical UE rate is below $0.1$ b/s/Hz (corresponding to 2 Mbps in a 20 MHz system bandwidth), the outage probability is about 0.49, 0.25, and 0.15, for $N=1,5,$ and 1$0$, respectively with FDMA/TDMA.

Comparing FDMA/TDMA and TDMA for the same $N>1$, it can be seen that TDMA is a better choice at low data rates. As stated above, at this value range it is the SIR outage probability that defines performance and TDMA is preferable as it is more robust to SIR outage events.  When higher data rates are considered, TDMA is penalized by the inability of concurrent UE transmissions and FDMA/TDMA becomes a better choice. In Fig. 4, this difference in performance is more clearly seen for $\tau=10$, which results in average $K$ and $K_0$ equal to 10 and 12.8, respectively. For this load and the values of $N$ considered, FDMA/TDMA utilizes all SCs with high probability, resulting in significantly larger interference compared to that achieved by TDMA which only allows for transmission on a single SC. However, for the same reason, performance of FDMA/TDMA is significantly better for higher rates as it provides much smaller delay than TDMA. Results for $\tau=1, 10$ show that the performance advantage of FDMA/TDMA in higher rates and of TDMA in lower rates still holds but is less pronounced.

\section{Optimal Number of Subchannels}

By simple examination of Fig. 4, it is understood that for a given minimum rate $r_0 > 0$, there is an $r_0$-dependent, optimal number  $N^*$  of SCs that minimizes rate outage probability which can be obtained by numerical search using the analytical expression of Proposition 1. Unfortunately, the highly non-linear dependence of $F_{R}(r_0)$ on $N$  does not allow for a closed form expression of $N^*$ that holds in the general case. However, the following proposition, valid under certain operational scenarios to be identified right after, provides some guidelines.
\newtheorem{proposition2}[proposition]{Proposition}
\begin{proposition2}
Under the assumption $\Pr\{L = 0\}=1$ (no time-sharing of SCs) and for any $\theta_0 \geq 0$, $r_0>0$, the optimal number $N^*$ of SCs  that minimizes $F_R(r_0)$ is lower bounded by 
\begin{equation} \label{eq:Nlb}
N^* \geq N^*_{\textrm{\emph{lb}}} \triangleq \max\left\{1,\left\lfloor \overline{R}_{\theta_0}/r_0 \right\rfloor\right\},
\end{equation}
\end{proposition2}

\begin{IEEEproof}
Setting $r=r_0$ and keeping only the term $L=0$ in (\ref{eq:F_R}), $F_R(\cdot)$ can be written as a function of $N$ as 
\begin{equation} \label{eq:F_R_LB}
\setlength{\nulldelimiterspace}{0pt}
F_{R}(N)=\left\{\begin{IEEEeqnarraybox}[\relax][c]{l's}
 F_{\textrm{SIR}}(\theta_0),& $N \leq \frac{\overline{R}_{\theta_0}}{r_0}$, \\
F_{\textrm{SIR}}\left(2^{r_0N}-1\right),& $N \geq \frac{\overline{R}_{\theta_0}}{r_0}$.
\end{IEEEeqnarraybox}\right.
\end{equation}
As shown in Sect. II. C, $F_{\textrm{SIR}}(\theta_0)$ is a decreasing function of $N$ for both TDMA and FDMA/TDMA, therefore, so is $F_R(r_0)$ for $N \leq \frac{\overline{R}_{\theta_0}}{r_0}$.
\end{IEEEproof}


As discussed in Sect. II. D, $L$ can be made arbitrarily small for both TDMA and FDMA/TDMA by increasing $\tau$, therefore, Proposition 2 holds asymptotically for $\tau \gg 1$, i.e., in an ultra dense AP deployment where $\Pr\{K_0=0\} \rightarrow 1$. However, FDMA/TDMA can also reduce delay by increasing $N$. It is easy to see that if, for a given $\tau$, $N^*_{\textrm{lb}}$ is larger than the minimum value of $N$ required for $\Pr\{K_0\leq N-1\} = 1$, i.e., no UEs sharing a SC, $N^*$ for FDMA/TDMA cannot be smaller than $N^*_{\textrm{lb}}$. These observations are summarized in the following corollary.

\newtheorem{corollary}{Corollary}
\begin{corollary}
The bound of (\ref{eq:Nlb}) holds for TDMA when $\tau$ is sufficiently large so that  $\Pr\{K_0=0\} \rightarrow 1$,  and for FDMA/TDMA when $\tau$ is sufficiently large so that $\Pr\{K_0\leq  N^*_{\textrm{lb}} - 1\} \rightarrow 1$.
\end{corollary}

Note that the lower bound of (\ref{eq:Nlb}) is inversely proportional to $r_0$ corresponding to the fact that, when lower user rates are considered, there is no need for large bandwidth utilization and the system can reduce interference by increasing $N$. For rates $r_0>\overline{R}_{\theta_0}/2$ the bound becomes trivial, i.e., equal to one, however, these rates may be of small interest in a practical setting as they lead to large rate outage probability, even with optimized $N$ and moderate load (see Fig. 4 and Sect. VI).

The following lemma guarantees that an arbitrarily large $N$ cannot provide any non-zero $r_0$, even though the SIR grows unbounded with $N$.

\newtheorem{lemma6}{Lemma}
\begin{lemma}
For $N\rightarrow \infty$, $F_R(r_0) \rightarrow 1$, for any $r_0 >0$.
\end{lemma}
\begin{IEEEproof}
It was shown in \cite{Andrews} that the mean of $\overline{R}$ is a strictly decreasing function of $N$. Since $0 \leq R \leq \overline{R}$, it follows that the mean of $R$ tends to 0 with increasing $N$, and applying Markov's inequality completes the proof.
\end{IEEEproof}

\section{Minimum Required Access Point Density for a Given Rate Outage Probability Constraint}

A common requirement in practical systems is to provide a minimum rate $r_0$ to their  subscribers with a specified, small outage probability $\epsilon>0$. As is clear from Fig. 4, these system requirements may be such that they cannot be satisfied for a certain $\tau$, even under optimized $N$. It is therefore necessary to operate in a greater $\tau$, i.e., increase AP density, and it is of interest to know the minimum value, $\tau_{\textrm{min}}$, that can provide the given requirements. As in the case of $N^*$, a closed form expression for $\tau_{\textrm{min}}$ can not be found in closed form for the general case and a two-dimensional numerical search (over $\tau$ and $N$) is necessary. However, under asymptotically small $\epsilon$, an upper bound of $\tau_{\textrm{min}}$ can be obtained for the case of FDMA/TDMA, as stated in the following proposition.
\newtheorem{proposition3}[proposition]{Proposition}
\begin{proposition3}
For FDMA/TDMA and asymptotically small $\epsilon$, the minimum value of $\tau$, $\tau_{\textrm{min}}$, that can support a UE rate $r_0$ with $F_{R}(r_0) \leq \epsilon$ under an SIR threshold $\theta_0$, is upper bounded as
\begin{equation} \label{eq:tmin}
\tau_{\textrm{min}} \leq  \left\{\begin{IEEEeqnarraybox}[\relax][c]{l's}
\frac{(1-\epsilon)\rho(\theta_0)}{\epsilon \left\lfloor \overline{R}_{\theta_0}/r_0 \right\rfloor},& $r_0 \leq \overline{R}_{\theta_0}$\\
\frac{(1-\epsilon)\rho(2^{r_0}-1)}{\epsilon}& $r_0 \geq \overline{R}_{\theta_0}$,
\end{IEEEeqnarraybox}\right.
\end{equation}

\end{proposition3}
\begin{IEEEproof} 
An upper bound on $\tau_{\textrm{min}}$ can be obtained by seeking the value of $\tau$ that provides the requested outage probability constraint with equality and under $N=N^*_{\textrm{lb}}$, as given in (\ref{eq:Nlb}), which is not guaranteed to be the optimal choice for $N$. In addition, only values of $\tau$ for which $N^*_{\textrm{lb}}$ results in $\Pr\{L=0\} = 1$ are considered, which effectively places a lower bound on the search space of $\tau$ that may be greater than $\tau_{\textrm{min}}$. Under these restrictions,
\setlength{\arraycolsep}{0.3em}
\begin{eqnarray} \label{eq:proof}
F_{R}(r_0) &\overset{(a)}{=}& \left\{\begin{IEEEeqnarraybox}[\relax][c]{l's}
F_{\textrm{SIR}}(\theta_0),& $r_0 \leq \overline{R}_{\theta_0}$,\\
F_{\textrm{SIR}}(2^{r_0} -1),& $r_0 \geq \overline{R}_{\theta_0}$,
\end{IEEEeqnarraybox}\right.\nonumber\\
&\overset{(b)}{=}& \left\{\begin{IEEEeqnarraybox}[\relax][c]{l's}
\frac{1}{1+\tau N^*_{\textrm{lb}}/\rho(\theta_0)}, & $r_0 \leq \overline{R}_{\theta_0}$,\\
\frac{1}{1+\tau/\rho(2^{r_0}-1)}, & $r_0 \geq \overline{R}_{\theta_0}$,
\end{IEEEeqnarraybox}\right.
\end{eqnarray}
where (a) follows from (\ref{eq:F_R}), (\ref{eq:F_R_L}) with $\Pr\{L=0\} = 1$ and (b) from (\ref{eq:FSIR}) and (\ref{eq:activ_prob}) with $\min\{K,N^*_{\textrm{lb}}\}=K, \forall K$. Setting (\ref{eq:proof}) equal to $\epsilon$ results in (\ref{eq:tmin}). Note that the asymptotically small $\epsilon$ guarantees that the bound of (\ref{eq:tmin}) is large enough such that $\Pr\{L=0\} \rightarrow 1$, i.e., it is within the restricted search space employed for its derivation.
\end{IEEEproof}
A simpler form of the bound can be obtained when asymptotic values of $r_0$ are considered as shown in the following proposition.
\newtheorem{proposition4}[proposition]{Proposition}
\begin{proposition4}
For asymptotically small or large values of $r_0$, the bound of (\ref{eq:tmin}) can be approximated by
\begin{equation} \label{eq:tminasympt}
\tau_{\textrm{min}} \leq  \left\{\begin{IEEEeqnarraybox}[\relax][c]{l's}
\frac{(1-\epsilon)\rho(\theta_0)}{\epsilon  \overline{R}_{\theta_0} }r_0,& $r_0 \ll \overline{R}_{\theta_0}$,\\
\frac{(1-\epsilon)2\pi}{\epsilon \alpha \sin(2\pi/\alpha)}2^{2r_0/\alpha},& $r_0 \gg \overline{R}_{\theta_0}$,
\end{IEEEeqnarraybox}\right.
\end{equation}
\end{proposition4}
\begin{IEEEproof}
The upper part of (\ref{eq:tminasympt}) can be obtained by noting that $1/\lfloor \overline{R}_{\theta_0}/r_0 \rfloor \approx r_0/\overline{R}_{\theta_0}$, for $r_0 \rightarrow 0$, whereas the lower part can be obtained by noting that $\rho(2^{r_0}-1)\approx \rho(2^{r_0}) = 2^{2r_0/\alpha} \int_{2^{-2r_0/\alpha}}^\infty 1/(1+u^{\alpha/2})du \approx 2^{2r_0/\alpha} \int_0^\infty 1/(1+u^{\alpha/2})du=2^{2r_0/\alpha}2\pi\sin(2\pi/\alpha)/\alpha$, for $r_0 \rightarrow \infty$
\end{IEEEproof}

Equation (\ref{eq:tminasympt}) clearly shows that the bound of $\tau_{\textrm{min}}$ grows linearly and exponentially with $r_0$, for asymptotically small and large $r_0$, respectively.

\begin{figure}
\centering
\resizebox{\columnwidth}{!}{\includegraphics{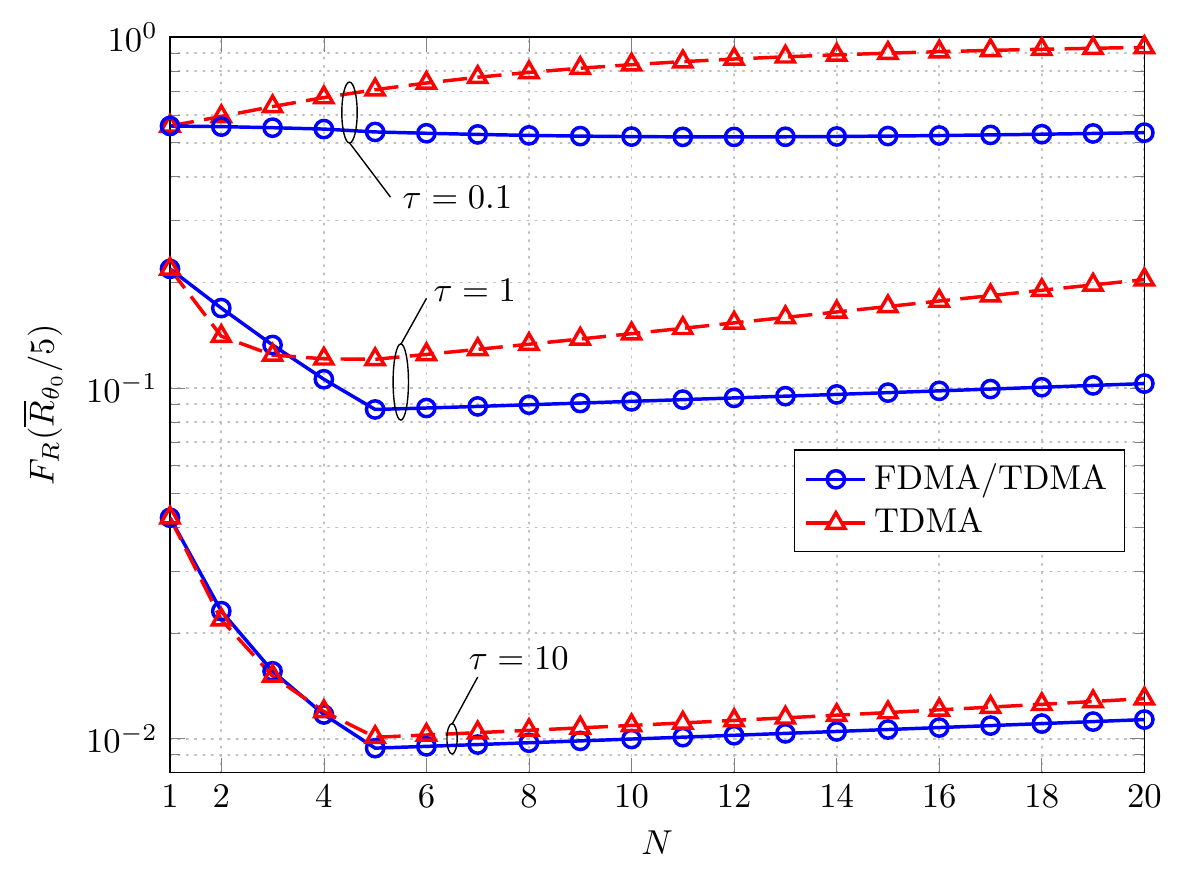}}
\caption{Dependence of $F_R(r_0)$ on $N$ ($r_0 = \overline{R}_{\theta_0}, \theta_0 = -6$ dB, $\alpha = 3$).}
\end{figure}

\section{Numerical Results and Discussion}
This section employs the analytical results obtained previously to examine various aspects of system design. In all cases the path loss exponent is set to $\alpha=3$ and, unless stated otherwise, the SIR threshold is set to $\theta_0 = -6\textrm{ dB}$ ($\overline{R}_{\theta_0}\approx 0.3233$), roughly corresponding to the operational SIR required by the minimum coding rate scheme of a real cellular system \cite{Piro}.

\begin{figure}
\centering
\resizebox{\columnwidth}{!}{\includegraphics{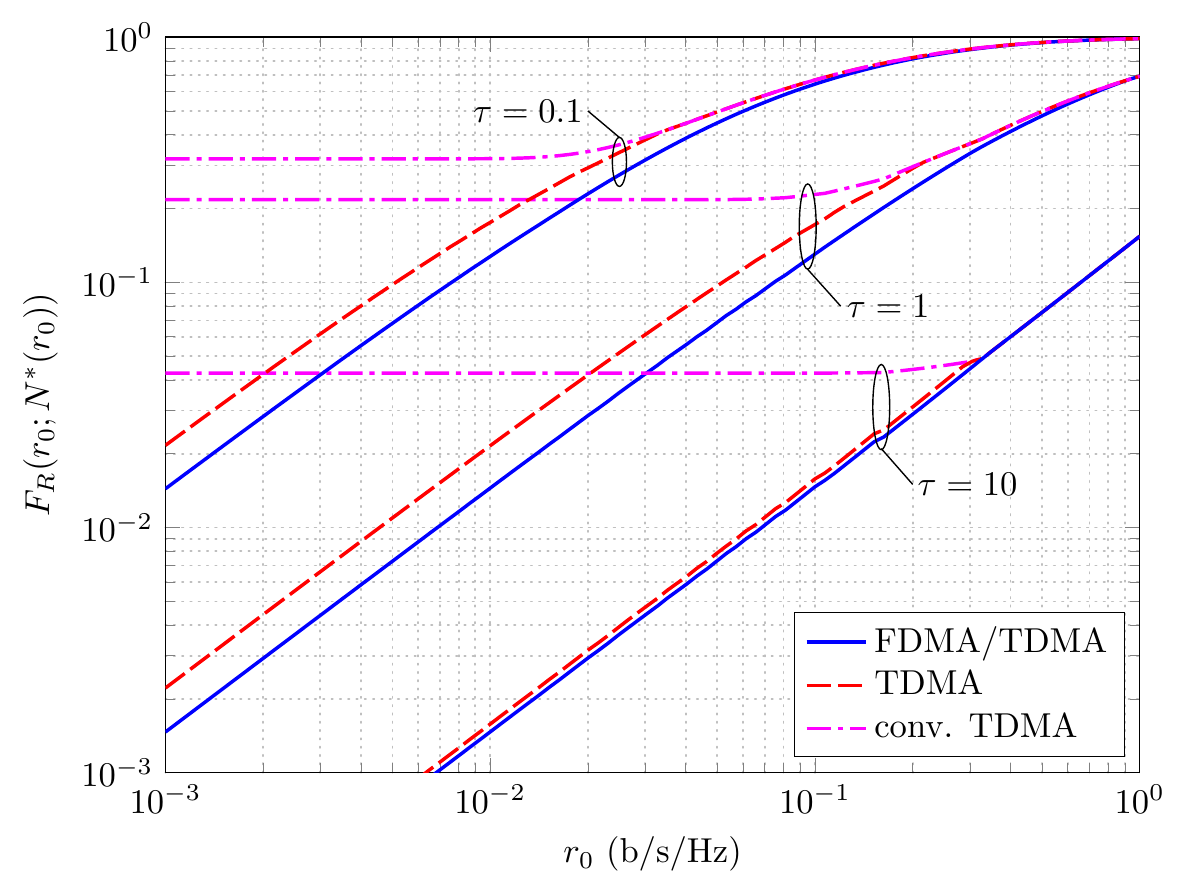}}
\caption{$F_{R}(r_0)$ with $N^*(r_0)$ ($\theta_0 = -6$ dB, $\alpha = 3$).}
\end{figure}

\emph{1) Optimal number of SCs:} Figure 5 shows $F_R(\overline{R}_{\theta_0}/5)$ as a function of $N$ for FDMA/TDMA and TDMA, and $\tau=0.1, 1$, and $10$. Note that by Proposition 2, $N^*$ is lower bounded by $N^*_{\textrm{lb}} = 5$ when $\Pr\{L=0\} = 1$. Consider first TDMA. It can be directly calculated that  $\Pr\{L=0\}=\Pr\{K_0=0\} \approx 0.0023, 0.3227, 0.8809$, for $\tau = 0.1, 1, 10$, respectively. Therefore, the operational conditions of Corollary 1 hold (approximately) only for the $\tau = 10$ case. It can be seen, that the bound is actually tight for that case, as $N^*=5$, whereas $N^*$ tends to one as smaller $\tau$ values are considered, i.e., $N^*_{\textrm{lb}}$ is a tight bound when $\tau \gg 1$ but is irrelevant for small $\tau$. Turning to the FDMA/TDMA case, $\Pr\{L=0\}=\Pr\{K_0 < N^*_{\textrm{lb}} -1\} \approx 0.1864, 0.9931, 1$, for $\tau = 0.1, 1, 10$, respectively, i.e., the operational conditions of Corollary 1 correspond to the cases of $\tau=1$ and $10$, with $N^*$ actually equal to $N^*_{\textrm{lb}}$. For $\tau=0.1$, $N^*=12$, i.e., $N^*_{\textrm{lb}}$ also servers as a lower bound in this case, albeit a loose one. However, note that performance gain with $N^*$ is only marginal compared to $N^*_{\textrm{lb}}$. These observations, along with extensive numerical experiments, suggest that setting $N=N^*_{\textrm{lb}}$ as per (\ref{eq:Nlb}) is a good practise for FDMA/TDMA as it either corresponds to the optimal value or provides performance close to optimal. For TDMA, setting  $N=N^*_{\textrm{lb}}$ for small $\tau$ may lead to considerable performance degradation.

\emph{2) Comparison of multiple access schemes with optimal $N$:} Figure 6 depicts the minimum rate outage probability provided by FDMA/TDMA and TDMA when the corresponding optimal $N$ for each rate $r_0$ is employed (found by numerical search). Note that these curves should not be confused as CDF curves since a different $N$ is employed for each rate. Performance of conventional TDMA is also shown. As can be seen, for small to moderate rates ($r_0 < \overline{R}_{\theta_0}$), optimal bandwidth partitioning provides significant benefits compared to conventional TDMA. FDMA/TDMA is shown to outperform TDMA in this regime as it exploits bandwidth more efficiently. For large rates ($r_0 \geq \overline{R}_{\theta_0}$) all schemes have the same performance as $N^*$ becomes one. It is safe to say that, under optimal $N$, FDMA/TDMA is preferable to TDMA as it provides at least as good performance with the added benefit of reduced delay that  is of importance under time-sensitive applications.

\emph{3) Effect of SIR threshold:} Figure 7 shows $F_R(r_0)$ as a function of SIR threshold $\theta_0$, for $r_0=0.1$, $\tau = 1$, and with $N$ optimized for each $\theta_0$ by numerical search. It can be seen that larger $\theta_0$ values result in degradation of performance for both FDMA/TDMA and TDMA, albeit much less severe than conventional TDMA. Also shown is the performance of FDMA/TDMA when $N$ is optimized assuming $\theta_0=0$, i.e., neglecting SIR outage. As expected, performance (significantly) degrades when the actual SIR threshold exceeds a certain value (about $-5$ dB in this case). Performance of TDMA assuming $\theta_0=0$ is not shown as it matches that of conventional TDMA. Similar behaviour is observed for other values of $r_0$, $\tau$ and $\alpha$. These results clearly illustrate the necessity of employing the SIR threshold in system analysis and design.

\begin{figure}
\centering
\resizebox{\columnwidth}{!}{\includegraphics{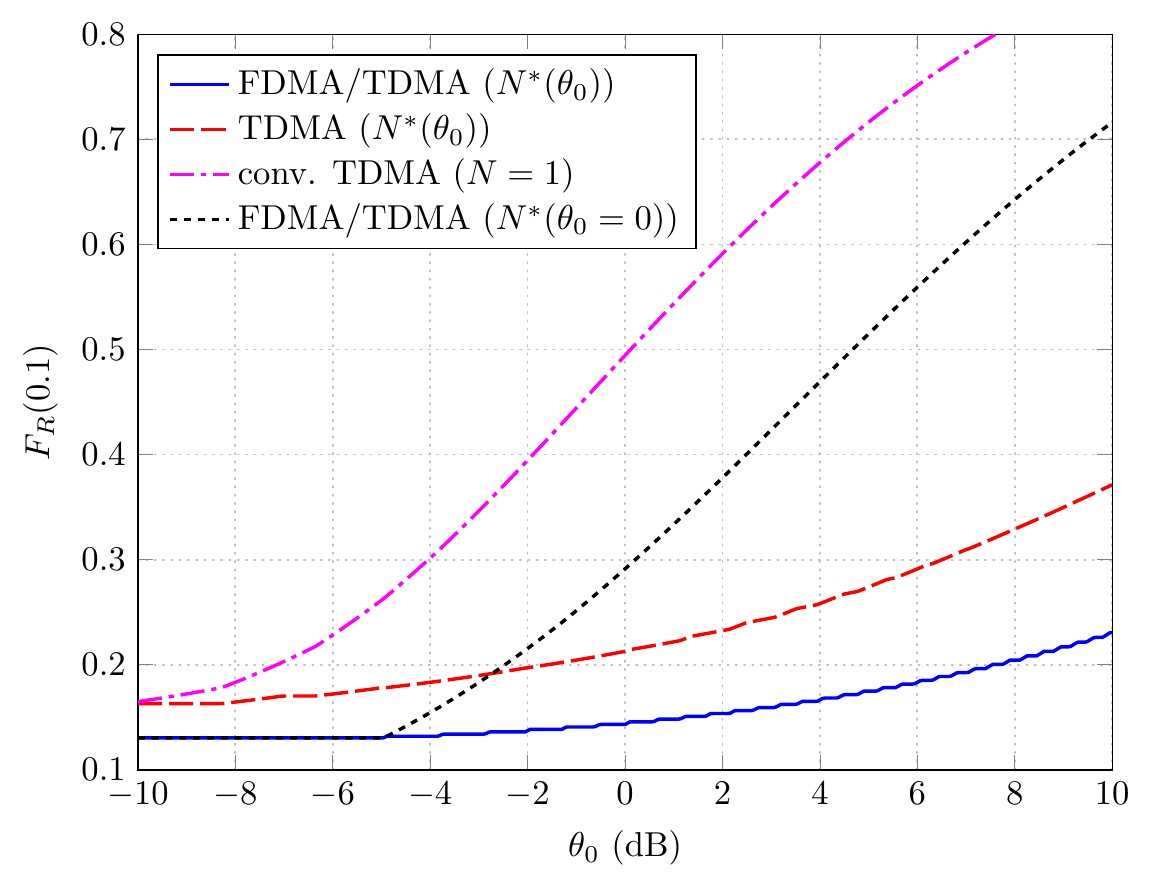}}
\caption{Dependence of $F_{R}(r_0)$ on $\theta_0$ ($r_0 = 0.1, \tau = 1, \alpha = 3$).}
\end{figure}

\emph{4) Minimum AP density:}
Figure 8 shows $\tau_{\textrm{min}}$  as a function of rate $r_0$, obtained by a two-dimensional numerical search over $\tau$ and $N$, for FDMA/TDMA, TDMA and conventional TDMA, under an outage constraint $F_R(r_0) \leq 0.1$. In addition, the asymptotic bounds of (\ref{eq:tminasympt}) are also shown. Note that, even though (\ref{eq:tminasympt}) is derived assuming asymptotically small $\epsilon$, it still provides a very good approximation of $\tau_{\textrm{min}}$ for this case. Specifically, $\tau_{\textrm{min}}$ of FDMA/TDMA exhibits the behavior predicted by (\ref{eq:tminasympt}), i.e., increases linearly and exponentially with $r_0$ for asymptotically small and large $r_0$ respectively. Performance of TDMA follows the same trend with FDMA/TDMA but results in about 1.5 times larger values of $\tau_{\textrm{min}}$ for small $r_0$. It is interesting to note that the very good correspondence of the numerical and analytical results for FDMA/TDMA implies that the  optimal system parameters ($\tau$ and $N$) for FDMA/TDMA are such that $\Pr\{L=0\}\approx 1$, i.e., there is small probability of sharing a SC. In contrast, TDMA achieves performance close to FDMA/TDMA with $\Pr\{L=0\}\approx 0$ for small $r_0$ ($\tau_{\textrm{min}}$). Conventional TDMA is clearly out of consideration for the small rate region as it significantly suffers from interference and the only mechanism to reduce it is by employing a large AP density. For rate values equal or greater than $\overline{R}_{\theta_0}$ all schemes coincide as the optimal value of $N$ turns out to be equal to one.

\begin{figure}
\centering
\resizebox{\columnwidth}{!}{\includegraphics{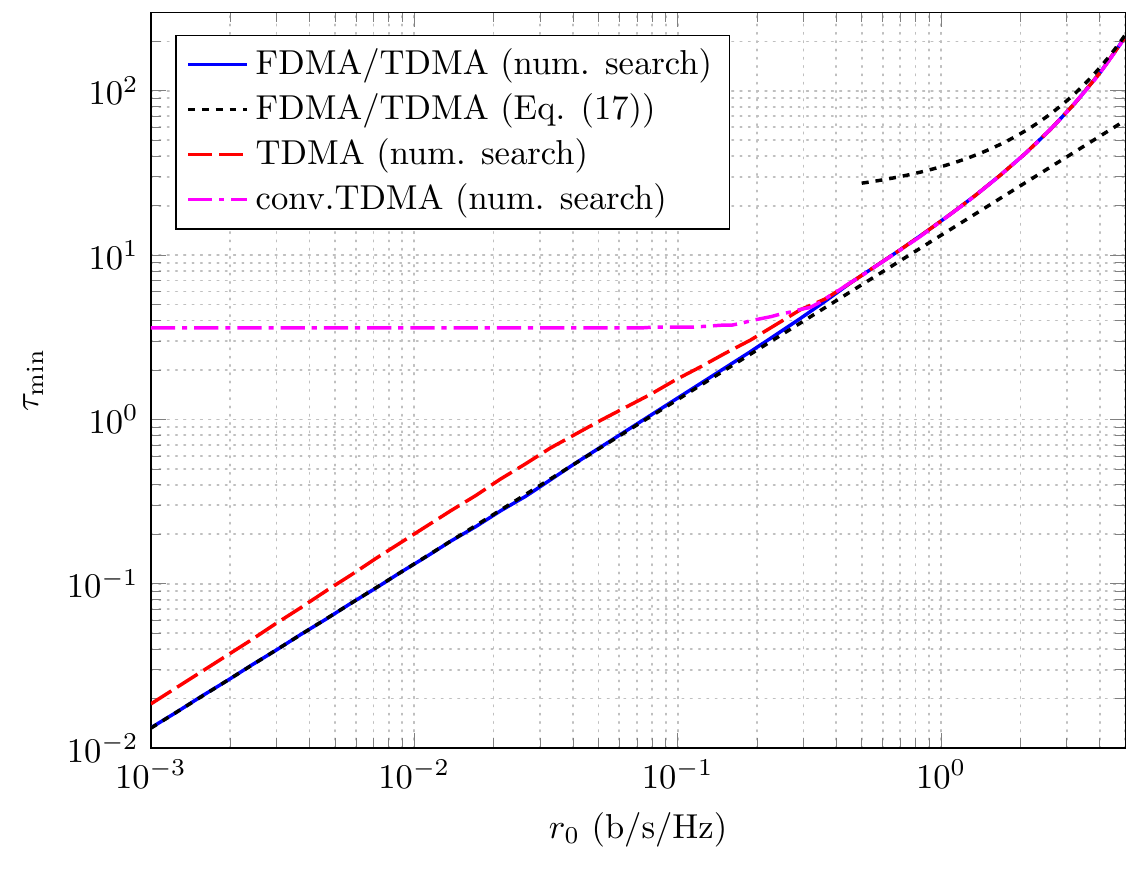}}
\caption{Minimum required $\tau$ for rate outage probability $\epsilon = 0.1$ ($\theta_0 = -6$ dB, $\alpha = 3$).}
\end{figure}

\section{Conclusion}
In this paper, system parameter selection, namely, number of bandwidth partitions and AP density was investigated for randomly deployed ultra dense wireless networks. The stochastic geometry framework from previous works was incorporated and the user rate distribution was derived analytically, taking into account the UE distribution, multiple access scheme and SIR outage. It was shown that performance depends critically on the number of bandwidth partitions and the way they are utilized by the multiple access scheme. The optimal number of partitions was tightly lower bounded under large AP density, showing that smaller bandwidth utilization is beneficial for interference reduction when small rates are considered. In addition, an upper bound on the minimum AP density required to provide an asymptotically small rate outage probability was obtained, that was shown to provide a very good estimate of the minimum density under moderate probability constraints. When the considered rates are small enough to allow for bandwidth partitioning, the minimum required density is smaller by orders of magnitude compared to the one provided by conventional TDMA.

\appendices
\section{Proof of Lemma 3}
Activity probability for TDMA is obtained by simply noting that $p = \Pr\{\textrm{transmission on SC 1}|K > 0\} \Pr\{K>0\}$. For the case of FDMA/TDMA, consider a random AP associated with $K$ indexed UEs and let $p(K)$ denote the probability of assigning at least one UE on SC 1. Clearly, $p(K)=1$ for $K \geq N$ and $p(K)=0$ for $K=0$. For the case $0<K<N$, define the mutually exclusive events $\mathcal{B}_m \triangleq$ $\{$$m$-th UE is assigned SC $1 \}, m=1,2,\ldots, K$. It is easy to see that 
\begin{equation} \label{eq:Pr_Bm}
\Pr\{\mathcal{B}_m\} = \frac{1}{N-(m-1)} \prod_{r=1}^{m-1}\left( 1 - \frac{1}{N-(r-1)}\right),
\end{equation}
and
\begin{equation} \label{eq:p_N_K}
p(K) = \sum_{m=1}^{K} \Pr\{\mathcal{B}_m\} = K/N, 0<K<N,
\end{equation}
where the last equality follows by simple algebra. Averaging $p(K)$ over $K$ results in the form of (\ref{eq:activ_prob}).

\section{Proof of (\ref{eq:F_R_L})}
Denoting the SIR outage event $\{\textrm{SIR} < \theta_0\}$ and its complement, $\{\textrm{SIR} \geq \theta_0\}$, as $\mathcal{O}$ and $\overline{\mathcal{O}}$, respectively, $F_{R}(r|L)$ can be written as
\setlength{\arraycolsep}{0.0em}
\begin{eqnarray}
F_{R}(r|L)&{}={}& F_{\textrm{SIR}}(\theta_0) F_{R}(r|L,\mathcal{O})\nonumber\\
&&{+}\:(1-F_{\textrm{SIR}}(\theta_0)) F_{R}(r|L,\overline{\mathcal{O}}).
\end{eqnarray}
\setlength{\arraycolsep}{5pt}
From (\ref{eq:4}), $R = 0$ conditioned on $\mathcal{O}$, therefore,
\begin{equation} \label{eq:F_R_L_a2}
F_{R}(r|L,\mathcal{O}) = 1, \forall r, L,
\end{equation}
whereas, conditioned on $\overline{\mathcal{O}}$,
\setlength{\arraycolsep}{0.0em}
\begin{eqnarray}
F_{R}(r|L,\overline{\mathcal{O}}) &{}={}& \Pr\{\overline{R}/(L+1) < r | \overline{\mathcal{O}}\}\nonumber\\
&{}={}& F_{\textrm{SIR}}(\tilde{\theta}|\textrm{SIR} \geq \theta_0)\nonumber\\
&{}={}& \left\{\begin{IEEEeqnarraybox}[\relax][c]{l's}
\frac{F_{\textrm{SIR}}(\tilde{\theta})-F_{\textrm{SIR}}(\theta_0)}{1-F_{\textrm{SIR}}(\theta_0)},& $\tilde{\theta} \geq \theta_0$\\
0,& $\tilde{\theta} < \theta_0,$
\end{IEEEeqnarraybox}\right.
\end{eqnarray}
where $\tilde{\theta} \triangleq 2^{rN(L+1)}-1$ and the last equality follows from basic probability theory and the continuity of $F_{\textrm{SIR}}(\theta)$. Combining (20)--(22) leads to (\ref{eq:F_R_L}) and application of the total probability theorem gives (\ref{eq:F_R}).

\end{document}